# Changing tools, changing habits, changing workflows: Recent evolutions of the interlibrary loan service at ULiège Library

Fabienne Prosmans and François Renaville

## Abstract

Over the last few years, the interlibrary loan (ILL) service of the University of Liège Library has evolved considerably, both in terms of habits and workflows. In this article, we will explain the main stages of this evolution: (1) first reduction in the number of ILL units (from eight to five) and involved operators (from 15 to 10) within the homemade ILL solution (2015); (2) use of the resource sharing (RS) functionality in the new Alma library management system (2015); (3) second reduction in the number of ILL units (from five to only one) and in involved operators (from 10 to six) (2018); (4) subscription to an international broker ILL system (RapidILL) for electronic and digital materials and its integration with Alma (2020); (5) project of peer-to-peer resource sharing for print materials between Alma instances of university and research libraries in Belgium (2022), and (temporary) free ILL service to all University users (2020–2022). The aim of these changes is to harmonise the practices of ILL operators, reduce the quantity of manual and administrative operations and tasks devoted to ILL and supply materials that do not belong to the library collections in a fairer, faster and more fluid way. While all of these changes have been implemented gradually over the years, not all of them have been deployed in a concerted manner among all stakeholders.

## Keywords

# Article

## Introduction

### *About ULiège Library*

The University of Liège (ULiège) was founded in 1817 by William I of the Netherlands, then King of the United Kingdom of the Netherlands, as it existed between 1815 and 1839. For a long time, the *Bibliothèque Générale* ('General Library') was the only real official university library. However, from the end of the 19th century, due to the expansion of university institutes throughout the city and the growing specialisation of fields of knowledge, certain departments and faculties started to develop their own collections and libraries ([Cuvelier et al., 2008](#)).

In 1954, the decision was taken to locate the whole university outside of the city, at the Sart Tilman hill. This led to a radical change in its documentary landscape in 1956, with the creation of specialised *Unités de Documentation* ('Documentation Units') that were close to users and therefore related to faculties or departments. All these units were characterised by a high degree of autonomy and independence. Very few services, such as cataloguing, were common and shared.

In the early 2000s, with the rise of digital technology, the university made the decision to modernise its libraries and documentation units, to reorganise them in depth and to group them into four, and subsequently five, major entities: Philosophy and Literature Library; Law, Economics, Management and Social Sciences Library; Science and Technology Library; Life Sciences Library; and Agronomy Library. "In addition to the much-needed modernisation and development of the electronic library, this restructuring aimed to improve and extend the service to users and to coordinate the acquisitions policies in a more efficient manner" ([ULiège Library, 2022](#); [Cuvelier et al., 2008](#)).

In January 2014, the members of the Library Board organised a 'greening' in order to define the main lines of reorganisation of the library and to review the library strategy in the context of the forthcoming bicentenary. They laid the foundations for various elements of reflection aimed at reviewing the strategy and missions, increasing collaboration between branches and services and reflecting on the necessary reconversion from a 'traditional collection library' to a new model where services are at the centre ('Library as a Service').

## ILL at ULiège Library, 2005–2015

From 2005 to 2015, eight library branches and 15 operators managed ILL services. We mainly used Impala and Subito[1] as partners for our resource sharing requests. Impala is used for both borrowing and lending requests, while Subito is only used for borrowing requests. It is important to note that the 15 operators were not working full-time on ILL services. Most worked less than 0.5 FTE, while some even less than 0.1 FTE.

Our ILS was Aleph, but we did not use its ILL module. Our interlibrary loan was managed within an in-house solution called MyDelivery, which was developed with APIs and integrated within the library website and the Primo discovery tool. A blank ILL request form was accessible in Primo MyAccount.

Interlibrary loan operators received an email informing them that a new request required processing. They then manually created the request in Impala or Subito and followed the evolution of the request in these brokers. MyDelivery was used for internal use only, e.g. for the creation, cancellation and management of requests, for invoicing and for some tracking features.

## Changes in ILL services

### 1) Reduction in the number of ILL units and operators

There were several reasons for which a reduction in the number of units and ILL operators was needed:
- The eight independent ILL units available were too many, and the smallest ones received very few requests. We wanted to avoid operators in these libraries having to process ILL requests only occasionally;
- Some libraries are not particularly far from each other. We wanted to free these libraries from the burden of providing an ILL service of their own if a close collaboration with another library located nearby could be established;
- We also wanted to avoid the risks of interruption of the ILL service for libraries with only one or two ILL operators. Even when libraries experienced no understaffing issues, some ILL units used to be closed for several weeks during holiday periods, for example;
- We migrated to Alma in early 2015. By reducing the number of ILL operators before the migration, we could consolidate the existing workflows between operators and reduce the number of operators to train in resource sharing (RS) in Alma.

---

1. Impala is the Belgian ILL platform for resource sharing while Subito is the German platform.

We decided to keep five ILL units and 10 operators (Fig. 1). At ULiège Library, interlibrary loan tasks do not occupy operators full-time since we estimated that the total ILL tasks represent 1.5 FTE.

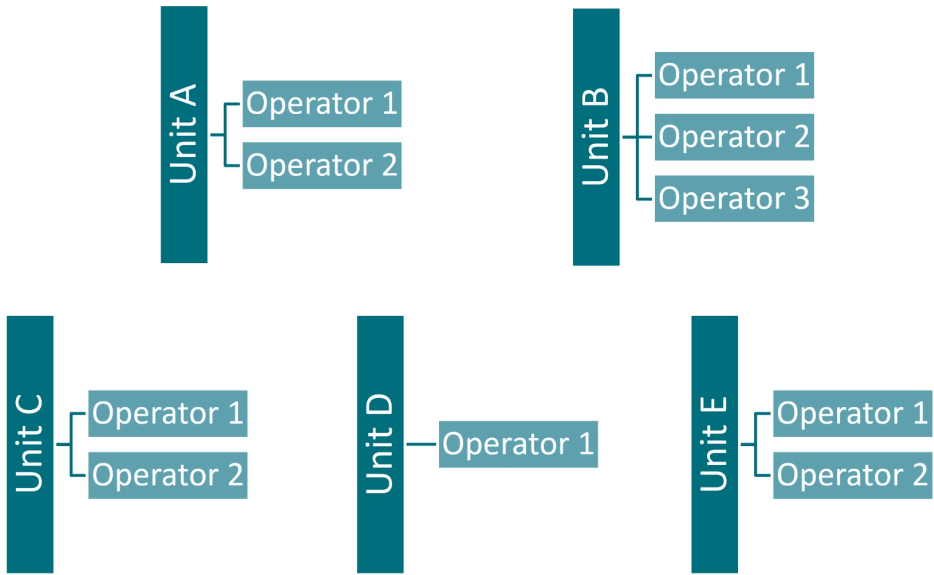

*Figure 1: New organisation after the first reduction (2015)*

After a few weeks, we noticed that this first reduction allowed us to simplify some tasks. Additionally, no negative impact on service quality was reported, and we did not receive any negative feedback from users.

Working with five ILL units allowed us to maintain significant freedom of organisation for each unit. However, we noticed that each library saw itself as an independent unit; there were very few interactions between operators and very little communication between units themselves. At that stage, the service was still strongly reliant on the library type: the unit associated with the Health Library used to only serve medicine students and members of the Faculty of Medicine and the University Hospital Centre, the unit associated with the Science Library limited its service to students and researchers in chemistry, physics, and astrophysics, etc.

This first reduction was accepted easily because of the migration to Alma: many organisational changes were expected, not only for ILL ([Bracke, 2012](#); [Renaville, 2018](#); [Brisbin et al., 2020](#)). Additionally, some ILL staff members saw this as a good opportunity to leave the service and to redirect their professional career within ULiège Library instead of changing their habits or following training sessions on resource sharing in Alma.

## 2) Moving to the Resource Sharing functionality in Alma

### Context

During the Alma implementation project, we decided to stop working with our in-house solution and to start fully using the resource sharing functionality in Alma. Our goal was to make the most of the new library system and to take advantage of Alma's various features.

Concretely, for borrowing requests, a book provided by a partner is temporarily located in one of the five resource sharing libraries and receives a temporary barcode. This allows the patron to loan the book. Letters such as the On Hold Shelf Letter, Courtesy Letter or Overdue Notice Letter are automatically managed and sent by the system, as these items are owned by the library. It is possible to define circulation rules (overdue fine, maximal renewal period, etc.) and to add resource sharing fees to the user account.

For lending requests, it is easier to know when a book is sent to a partner for resource sharing purposes. Previously, with the in-house solution, operators had to lend the book in their own name. Moreover, with Alma, the borrowing partner receives the Borrower Overdue Email Letter when the due date is reached.

Finally, it is easy to retrieve statistics from Alma Analytics such as the number of borrowing requests by user group or the number of borrowing or lending requests by partner or material type.

This allowed us to stop using our home-made solution MyDelivery, but we still had to use our main partners (Impala and Subito) for delivery.

### What we did at ULiège Library

Each of our five ILL units was associated in Alma to a library for which the option 'Is resource sharing library' had been checked. In Alma, libraries "within an institution or campus may be configured to have relationships where they enable patrons to check in or check out resources at another location, send items back and forth, or acquire (purchase) items on behalf of each other. If a library is configured to do this for libraries at other institutions (and not only within the institution), it is known as a *resource sharing library*" ([Ex Libris, n.d.-a](#)).

We assigned the most appropriate resource sharing library to each of our 40,000 users (students and faculty and staff members) based on their field of research or studies.

Then, we created two specific locations for each Resource Sharing Library in Alma. The first location was for borrowing requests. A physical document that comes from a partner is temporarily assigned to this location; as mentioned above, the document receives a temporary barcode which allows the user to loan the document. The other location was for lending requests; documents that go to a partner 'leave' their original location and are temporarily located in this new location.

We also created two circulation rules by library, one for borrowing and one for lending requests. Libraries did not always define the same rules. Some libraries allowed readers to borrow materials from a partner, while others only accepted consultation in the reading room.

Finally, we had to create our partners for resource sharing in Alma. At this stage, all our partners were of the email profile type.

There are three ways to create borrowing requests in Alma at ULiège Library (Fig. 3):

1. First, for a resource not available at the University of Liège, the patron can use the View It or Get It tab in Primo or a database. In this case, we used the Ex Libris's resource sharing form where metadata are automatically taken from the bibliographic record. Once the form is submitted, a borrowing request is created in Alma, and interlibrary loan operators can find it in their task list;
2. Users can use a blank form for their ILL requests. Since we wanted a more customised request form than the one proposed by Ex Libris by default, we set up a new blank form (Fig. 2) and integrated it with the Alma APIs (De Groof, 2017). Once the form is submitted by the patron, the Alma API creates a borrowing request in the system, and ILL operators can start working on it;
3. Patrons can send an email to the ILL unit with their request details. Operators then manually create the request in Alma and start the fulfilment process.

*Figure 2: In-house blank form*

Compared to the Ex Libris default resource sharing blank form used at the time, we found that ours had a lot of advantages:
- It can be used for borrowing requests created by registered and non-registered

users such as university services or third-party services. It can also be used for lending requests created by other libraries or partners;
- The blank form is interactive and adapts according to the selections made (requested material, etc.), and only the necessary fields are displayed;
- If the user enters the DOI or PMID, the fields for bibliographic details are automatically filled in;[2]
- The journal title field is interfaced with a locally managed journal database containing over 45,000 journals, which makes it possible to obtain the standardised journal title and the ISSN in the ILL requests;
- The 'delivery and payment' fields are fully configurable according to the status or user group of the requester;
- In case of physical delivery, it is mandatory to select a pickup location from the drop-down list. No pickup location field is displayed for digital delivery.[3]

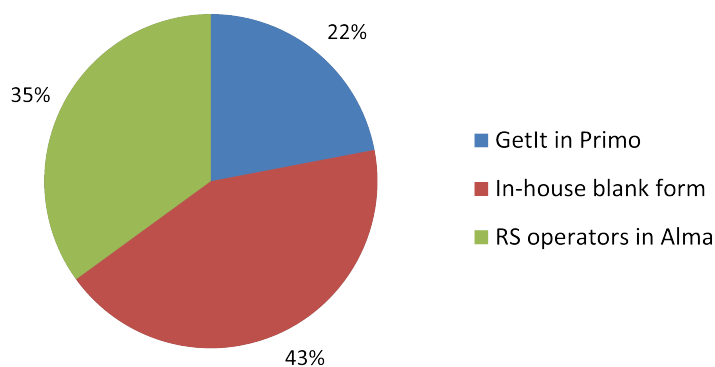

*Figure 3: Origin of borrowing requests (2017)*

In February 2015, when ULiège Library went live with Alma, we stopped using MyDelivery and exclusively worked with our five resource sharing libraries in Alma.

## 3) Simplification of the ILL backend

Working with five ILL libraries and 10 operators represented an improvement, but that number was still too high. Therefore, in autumn 2017, the Library Board decided to reor-ganise and centralise the interlibrary loan service and to create a new single entity. From January 2018, the new entity would virtually group all ILL operators together, who would work as a team and serve all library branches and patrons.

---

2. This feature is now also available in the Primo default form.
3. Currently, the pickup location is still proposed in the Alma/Primo blank form. See: Hide Pickup Location on Blank ILL Article/Book Chapter Request Form https://ideas.exlibrisgroup.com/forums/308173-alma/suggestions/46394707-hide-pickup-location-on-blank-ill-article-book-cha.

We deactivated the five existing resource sharing libraries and decided to use the default resource sharing library in Alma. This library is only used for resource sharing.

Due to the new organisation and the pooling of human resources, it was no longer necessary to spread the work across 10 librarians, so we also reduced the number of operators to six (Fig. 4).

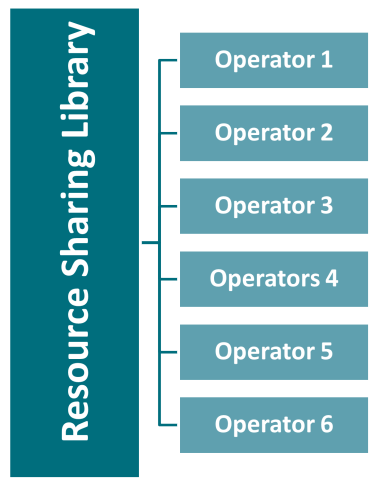

*Figure 4: Virtual Resource Sharing Library – New organisation after the second reduction (2018)*

Although ILL operators do not physically work at the same place, the resource sharing library has a physical address; this is important for physical items provided by partners. Thanks to a shared calendar, operators complete their resource sharing tasks in turn. They only spend a few hours per week on these tasks, and they complete these tasks in the library where they are employed.

Thus far, we have experienced various advantages of using only one resource sharing library:
- The same resource sharing library is now automatically allocated to all users. Previously, the resource sharing library allocated to users was based on each user's field of research or studies. Not all users had the most appropriate resource sharing library assigned to them in their user record, and some users had no assigned resource sharing library at all.
- This centralisation has reduced the number of transits of physical documents for borrowing requests.

- ILL operators collaborate more closely now as they are all members of the same virtual team.
- There is more fluency in ILL delivery. All ILL operators complete their RS tasks in turn. Additionally, in the past, there was the possibility of all ILL operators being out of the office at the same time (because of sickness, holiday, etc.), thus interrupting the ILL service in a library. Now, the probability that all six operators are absent at the same time is close to zero, and a substitute can easily be found if needed.
- The practices of the resource sharing operators have been strongly harmonised. Previously, there were various differences in workflows. For example, some RS libraries allowed the loan of books (provided by partners) to local researchers, while others only allowed consultation in the reading room. Now, all operators follow the same procedure, and there is one single circulation rule for borrowing requests and one single rule for lending requests.

However, we encountered some difficulties during this transition from five resource sharing libraries to one. The interlibrary loan service was the preserve of some operators. The transition involved more transparency across ULiège Library and the pooling of human resources, and costs and profits became shared at the top level. The change was not easily accepted by some colleagues: some operators had concerns regarding having more work, while others regarding having too little work. Some operators worried about having to process requests related to less familiar disciplines. Some operators also regretted the loss of conviviality due to the automation of certain tasks, such as automatic emails sent from Alma. Finally, since the Library Board decided to reduce the number of libraries from five to one at the end of November 2017, we only had a few weeks to change the workflows to be ready for early 2018. This was not easily accepted by some operators, as several thought the change had been applied too fast ([Prosmans & Renaville, 2018](#)).

## *4) Moving forward with RapidILL for digital requests*

For almost 30 years, ULiège Library had used the Impala system as its main ILL platform for managing its outgoing and incoming requests. Developed by the University of Antwerp Library in 1990, Impala was chosen in 1992 as the document delivery platform for interlibrary loan by the national conference of chief librarians of Belgian universities ([Corthouts et al., 2011](#)). Impala is accessible to all types of libraries (academic, research and public) and can handle different types of documents. Journal articles and physical books are mainly provided; Impala does not handle the provision of e-books. Of the 177,458 requests processed in 2000, the year with the highest number of requests, 134,556 requests came from university libraries (23,660 for books and 110,896 for articles) – which, however, only represent about 10% of Impala's membership ([Corthouts et al., 2011](#)). This shows the important use of the platform among Belgian university libraries.

However, despite the reliability of the tool and the fact that the Impala solution was still evolving ([Saerens, 2014](#)), ULiège Library representatives disliked the fact that there was no integration between the Impala and Alma systems. Until early 2015, with the use of the in-house MyDelivery application, incoming requests from Impala partners were managed exclusively within Impala. With the switch to Alma and in order to work in the most integrated way possible with the new system, it emerged that ILL staff often had to enter data twice: requests via Impala (lending requests) were entered manually into Alma, whereas direct requests from users (borrowing requests) using the Primo form or the blank form had to be re-entered by the operators (copy and paste) into Impala in order to be fulfilled. This was particularly inefficient, and there was a need for simplification. This situation was also present between Alma and Subito, but with the latter being a more occasional supplier compared with Impala, the cumbersome nature of the need for double data entry was less noticeable.

In spring 2020, the COVID-19 pandemic had a strong impact on library services and accessibility, and many libraries tried to reinvent themselves or innovate where possible ([Whitfield et al., 2020](#); [Ashiq et al., 2022](#); [Skalski, 2023](#)). In March, Ex Libris, which had acquired the RapidILL solution in 2019 ([Ex Libris, 2019](#); [Baeyens, 2023](#)), offered its customers the opportunity to join the new COVID-19 pod to test the resource sharing solution and participate in the collective effort to address the pandemic ([Veinstein, 2020](#)). Interested libraries could simply ask to join the COVID-19 pod to submit requests for articles and book chapters (digital delivery), with the RapidILL client community fulfilling their requests as much as possible.

ULiège Library saw this as an excellent opportunity to test a new tool at a time when it was needed:
- Due to the lockdown that started on 18 March 2020 in Belgium, the library had to close its doors and limit its services and activities to anything that could be executed remotely. The ILL service was therefore heavily impacted.
- Most of the libraries we worked with in the Impala network also temporarily ceased their activities, and it was difficult, if not impossible, for the physical provision to meet the ILL requests of our users.

Participation in the COVID-19 pod was therefore timely for us, and ULiège Library joined the pod in a trial. The results of this trial proved to be conclusive for our library.

Between 23 April and 31 August 2020, approximately 200 requests sent to RapidILL were satisfied (a satisfaction rate of 95%). Delivery time ranged from a few hours to, exceptionally, a few days – e.g. when requests were sent on a Friday at the end of the day.

Regarding the learning curve, the ILL team members expressed their satisfaction with the use of RapidILL. They found that this new solution provided added value, especially when requests for items were more difficult to meet using Impala.

In addition, unlike Impala, RapidILL can interface with Alma to avoid double data entry, which saves time considerably for staff. Although this integration was not tested in the trial, it was of great interest to the staff. This Alma/RapidILL integration not only allows users' requests to be transferred directly to RapidILL but also, when these requests are fulfilled by a RapidILL partner, to mark the request as completed in Alma and to send, from Alma, a delivery notification letter with a download link directly to the user. The automation process is such that a student or researcher can now submit a request for an article via the form and receive the requested copy a few hours later, without any intervention from ILL staff.

RapidILL's business model is based on a fixed annual fee set out in a contract. The number of requests processed by RapidILL is therefore irrelevant – unlike in Impala or Subito, where there is a cost per use. However, there is an obligation of means to be provided since RapidILL's clients are both borrowers and lenders ([Delaney & Richins, 2012](#)).

Finally, the predictable cost of a RapidILL subscription also allows libraries to consider certain issues and services differently. For example, it may become easier to forego certain non-essential journal subscriptions if their articles can quickly and easily be supplied through RapidILL. Similarly, offering a free ILL service to students and researchers becomes an easier option to consider ([Renaville & Prosmans, 2023](#)).

In view of these positive elements following the trial and analysis of the possibilities, ULiège Library therefore decided to subscribe to the RapidILL solution in September 2020. While it had initially seemed interesting to us to integrate a French-speaking pod or a pod with other Belgian libraries which would have also subscribed to RapidILL, this option quickly proved pointless because the default pods which we had joined were sufficient for our needs. Moreover, one must keep in mind that the more pods a library joins, the more likely they are going to be asked to supply other pod members with materials ([Baeyens, 2023](#)). A balance must therefore be struck between the two.

Although no French or Belgian pods were created, we were able to benefit from the arrival of several important new partners for us in the RapidILL community:
- Belgian universities: KU Leuven, Ghent University, Free University of Brussels (ULB);
- French universities: Université Clermont Auvergne;
- Swiss universities: BCU Lausanne, SLSP consortium (via Rapido).

While RapidILL was initially an additional supplier for article and book chapter requests, it became the preferred partner (by default) from November 2020 onwards thanks to the activation of processing automation between Alma and RapidILL (Fig. 5).

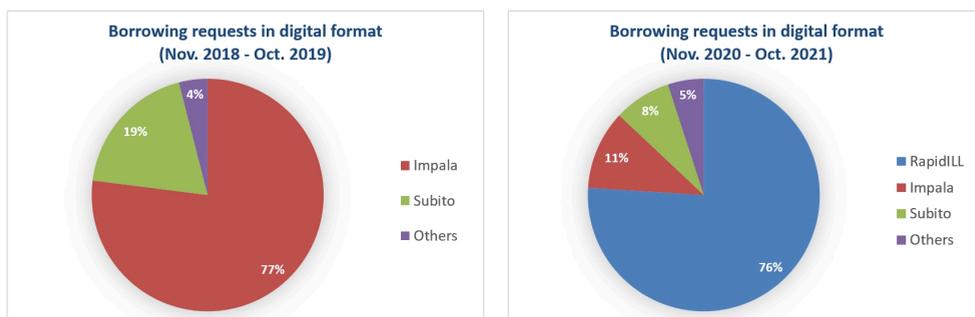

*Figure 5: Distribution by partner for digital delivery*

After a few months of use, we were able to establish the following advantages of working with RapidILL:

- Full and easy integration with Alma: Our ILL operators only sign in to RapidILL to access statistics or check potentially problematic cases;
- Fast delivery of requested documents: 10 hours on average for 2021 and 12 hours for 2022 (Table 1);
- Known annual cost (high borrowing usage has no financial impact);
- The growth of the Rapido solution (fully integrated in Alma itself), which benefits the RapidILL customer community as every new Rapido customer also joins the general COSMO pod (Ex Libris, n.d.-c);
- The fact that it is a global community and, therefore, that RapidILL libraries' collections are richer than Impala's historical partners (Table 2);
- The fact that it is not necessary to be an Alma customer to use RapidILL. For example, in 2021, the Internet Archive became a provider within RapidILL (Kahle & Coelho, 2021).

|  | **2021** | **2022** |
| --- | --- | --- |
| ULiège Borrowing Requests | 2,267 | 2,391 |
| System Average Borrowing Requests | 2,947 | 2,898 |
| ULiège Borrowing Filled | 2,009 | 2,187 |
| ULiège Borrowing Unfilled | 199 | 146 |
| ULiège % Filled | 89% | 91% |
| System Average % Filled | 95% | 95% |
| ULiège Average Filled TAT (Hours) | 10 | 12 |
| System Average Filled TAT (Hours) | 13 | 13 |

*Table 1: RapidILL borrowing statistics for 2021 and 2022*

| Lending Institution | Number of Requests |
|---|---|
| Internet Archive | 283 |
| Tufts Univ., Hirsh Health Sciences Lib. | 139 |
| Université de Clermont-Ferrand | 102 |
| Université Libre de Bruxelles | 97 |
| Rowan University | 90 |
| University of Pennsylvania | 86 |
| Olivet Nazarene University | 83 |
| Univ. of MD Baltimore Health Sciences Library | 83 |
| Ghent University Library | 69 |
| University of Chicago | 64 |
| Binghamton University | 60 |
| Oregon Health and Science University | 53 |
| Indiana Univ./Purdue Univ. Indianapolis | 51 |
| University of Arkansas Libraries | 49 |
| KU Leuven LIBIS | 48 |

*Table 2: Top 15 of our RapidILL lending suppliers for digital delivery in 2021 and 2022*

On the negative side, there is currently no priority for electronic holdings over print holdings of a lender library in RapidILL. Presently, format is not a parameter in the algorithm used to select the most appropriate partner. If it were possible for electronic holdings to take priority over physical holdings, it would help institutions to avoid digitising print materials (a procedure that requires time and human resources) if others have an electronic version that can be provided more easily and quickly. This is, however, a request for improvement that has been introduced and is supported by the customer community.[4]

## 5) Peer-to-peer resource sharing between Alma instances for physical items

While digital delivery of articles and book chapters was now mainly conducted via RapidILL, Impala remained the preferred solution for physical delivery (almost exclu-sively books). Here, too, however, there was a need for interfacing: ULiège Library's ILL staff still had to manually re-enter borrowing requests from our users into Impala.

---

4. Electronic holdings should take priority over physical holdings https://ideas.exlibrisgroup.com/forums/935109-rapidill/suggestions/43633749-electronic-holdings-should-take-priority-over-phys.

In 2021, [Ex Libris (n.d.-b)](#) introduced the Resource Sharing Directory:

> The directory is a central place with up-to-date information about resource sharing libraries in Alma. Libraries opt-in and define themselves in the directory with a few simple steps. Being in the directory will allow other libraries to find your institution and create a peer-to-peer resource sharing relationship with you simply and with less back and forth communication over technical connection details.

Libraries in the Alma Resource Sharing Directory are grouped into regional pods. In this way, they can easily create peer-to-peer relationship for physical supply. ULiège Library joined the Resource Sharing Directory by the end of 2021. In Belgium, the Free University of Brussels (ULB) and the European Commission Library also joined the Directory.

Partnering within the Directory brings several advantages:

- It helps to avoid filing encoding and tracking borrowing requests for physical items in the Belgian ILL platform Impala.
- Several steps of the workflow can be made automatic in Alma:
  - Sending a borrowing request from Alma to the partners of the Directory can be automatic;
  - With rules created by the library, partners from the Directory can easily be automatically assigned to requests in physical format;
- It allows a reduction in the cost of the ILL service: each request made via Impala is charged at the broker system level (cost per use). Using the Resource Sharing Directory is free of charge and no borrowing costs are expected between partners. Additionally, if the costs of the service decrease, this represents an additional argument in favour of a free ILL service for the benefit of our students and researchers.

Working with the Free University of Brussels (ULB) Library has been successful so far. In the first two months of 2023, out of a total of 78 successful physical borrowing requests, 28 were delivered by the Free University of Brussels and 34 via the Impala network. Previously, all requests completed by ULB would have been processed within Impala.

We hope that we can now expand this P2P experience to other partners in Belgium, such as KU Leuven or libraries from European institutions.

## 6) Free ILL service for our patrons

In 2018, the members of the Library Board decided to conduct a feasibility study on the possibility of offering a free ILL service to the university community. The study started in 2019 and was not completed until 2022, partly due to the pandemic. However, the pandemic offered some interesting insights into the study. To mitigate the impact of the national lockdown in March 2020 on library services, it was decided that the ILL service would become free for students, faculty and staff during the pandemic. As a consequence, and unexpectedly, the lockdown allowed for a real pilot phase for the free service, with consistent usage data, in parallel to the study itself ([Renaville, 2020](); [Renaville & Prosmans, 2022](); [Renaville & Prosmans, 2023]()).

Feasibility study

In order to avoid rushing into this project and repeating the mistakes made a few years ago when the five ILL units were merged too fast into one service (see above), it was decided to take time for analysis. The study paid particular attention to:
- The internal context;
- The external context;
- The stakeholder analysis;
- A strengths, weaknesses, opportunities and threats (SWOT) analysis.

*Internal context*

This part of the study focused on the current organisation of the ILL service, the history of the service and the various recent changes (see above), including the tools used so far, and several statistical data going back to 2018 (outgoing requests by format and faculty, profiles of requesters, suppliers, costs and revenues).

*External context*

In addition to a literature review, attention was also paid to the national context, and a comparative analysis with other Belgian universities was conducted: Did other universities offer a free service? If so, under what conditions and for whom? Finally, two conjunctural elements were added: the COVID-19 crisis which closed libraries and thus temporarily prevented any physical supply, increasing in parallel the digital demands, and the trial of the RapidILL solution, becoming in itself an internal influencing factor.

*Stakeholder analysis*

The objective of this section was to identify the different stakeholders impacted by the project:
- Primary stakeholders: most affected, either positively or negatively;
- Secondary stakeholders: indirectly affected;
- Tertiary stakeholders: least impacted.

A power-interest grid was also helpful to identify, and eventually contain, the stakeholders' interests, any potential risks or misunderstandings ('me issues'), any mechanisms that positively influence other stakeholders and the project and, finally, any negative stakeholders as well as their adverse effects on the project. An important aspect of the work consisted in individual interviews with eight key stakeholders, i.e. all ILL staff members and those responsible for billing the service. Surprisingly, the stakeholder analysis showed that all ILL operators were in favour of a completely free ILL service for students, and the only ones (slightly) against a free service were was among the sponsor themselves (the Library Board, in 2021).

*SWOT Analysis*

All these elements from the internal and external factor analyses and the stakeholder analysis were then synthesised in a SWOT analysis, which, as the acronym suggests, was a compilation of the project's strengths, weaknesses, opportunities and threats

## Pilot phase during COVID-19

The pilot phase was conducted from spring 2020 (first lockdown) to spring 2022, which is an exceptional length of time to have solid statistics and usage data, but conversely, so long that reverting to a paid service would have been difficult and would have required evidence in the form of consistent data and arguments. In addition to the statistics, the ILL operators also had two years' experience in a free context.

*Usage data*

While the provision of physical documents decreased between 2019 and 2020 due to the temporary closure of some ILL services, there was a very significant increase in digital requests from 2020 onwards, mainly due to the free service provision (Fig. 6). Unsurpris-ingly, it was mostly university staff members (x2 between 2020 and 2021) and graduate students (x3 between 2020 and 2021) who benefitted the most from the service. Although undergraduates are moderate users of the service in absolute terms, they used it five times more between 2020 and 2021 (Fig. 7).

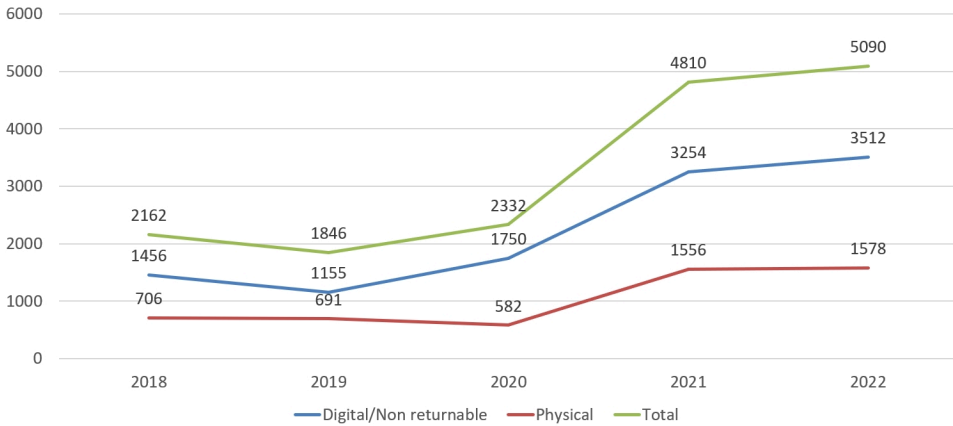

Figure 6: Completed borrowing requests (2018-2022)

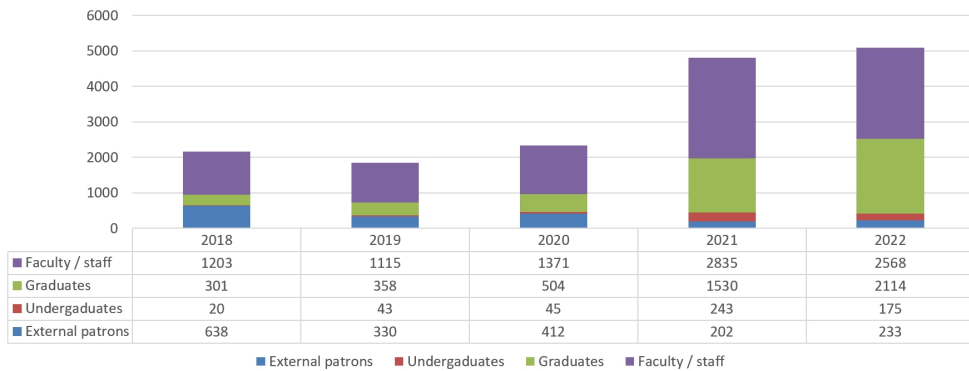

Figure 7: Completed borrowing requests by patron type (2018-2022)

*First reactions among users and ILL staff*

Not surprisingly, feedback from students on the free service was positive, and the increase in requests suggested that a need had been met.

Among ILL operators, it was appreciated that there was no longer any administrative burden (ILL fees to be charged to students' accounts, invoices to be managed, follow-up of payments, etc.).

There were some rare abuses from users in physical requests, but most striking was probably the fact that for the supply of books outside Belgium, postal charges were now borne exclusively by the library, which increased the cost of the service. However, due to the subscription to RapidILL and its integration with Alma (see above), the significant increase in the number of requests did not impact the library's finances or the workload of the staff.

Recommendations to the Library Board Committee

The feasibility study, consolidated by a two-year pilot phase, recommended to the Library Board to consider that free ILL should be a priority within the Library as a Service model and that it should be maintained for the ULiège community:
- With no limit for digital articles and chapters delivered via RapidILL;
- With an increase budget for postal charges;
- With no symbolic contribution to the costs.

In parallel to these measures, Alma P2P for physical delivery should be encouraged. The study also recommended some practical safeguards to prevent abuses and avoid uncontrolled costs, including purchasing a print copy or an e-book version when necessary ([Van den Avijle & Maggiore, 2023](#)), and to raise awareness among patrons that there are always hidden costs for the university and the library.

# Conclusion

The ILL service of the University of Liège Library has undergone major changes for seven years: changes in habits with the reduction of the number involved operators and ILL units and the reorganization of the team, changes of tools (from a homemade solution to Alma/RapidILL), and changes in workflows for more integration and automation.

Thanks to these changes, it was observed that the professionalisation and involvement of the staff have increased and that processes have become more highly standardised. A smaller number of people does not imply a lower-quality service. On the contrary, delivery is even more efficient as the service is rarely closed and supply is even faster. ILL operators were not necessarily against the changes, but they must be engaged in the evolution of the service because they need to understand what is taking place and must want to be involved in the thinking, decisions and changes in their work. Because change management takes time and preparation, it is important that managers carefully consider their priorities in implementing changes and take it step by step. From an institutional point of view, peer-to-peer resource sharing has led to new collaborations with partners. Finally, if costs remain under control, a free ILL service is possible.

# About the Authors

Fabienne Prosmans, Fulfilment and ILL Manager
Université de Liège (University of Liège)
https://orcid.org/0000-0002-1408-5207

Fabienne Prosmans (1972) studied mathematics at the University of Liège and obtained a PhD at the University of Paris Nord. Since 2005, she has been working at ULiège Library as a subject librarian in the fields of mathematics and applied sciences. She has been the fulfilment and ILL manager at ULiège Library since 2015 and is now coordinating the "User Services" unit.

François Renaville, Head of Library Systems
Université de Liège (University of Liège)
https://orcid.org/0000-0003-1453-1040

François Renaville (1976) studied linguistics, literature, and translation. After a two-year experience as a teacher in Finland, he joined the University of Liège Library as a subject librarian. He has been working on library systems since 2005. Since 2022, he has been member of the IGeLU (International Group of Ex Libris Users) Steering Committee and has been coordinating the "Systems & Data" unit at ULiège Library. He is interested in discoverability, integrations, delivery and user services. On a private level, he is a great coffee, chocolate and penguin lover.

# Beyond the Library Collections

*Proceedings of the 2022 Erasmus Staff Training Week at ULiège Library*

François Renaville and Fabienne Prosmans (Eds.)

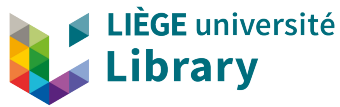

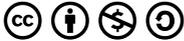

# Contents